# Communications Law

2010

## The ABC of Digital Business Ecosystems

Jo Stanley
Lucy Cavendish College
University of Cambridge
js731@cam.ac.uk

Gerard Briscoe
Department of Media and Communications
London School of Economics and Political Science
g.briscoe@lse.ac.uk

**Subject:** Legal foundations for network-based economies

**Abstract**

The European Commission has the power to inspire, initiate and sponsor huge transnational projects to an extent impossible for most other entities. These projects can address universal themes and develop well-being models that are valuable across a diversity of societies and economies.

It is a universal fact that SMEs in all countries provide a substantial proportion of total employment, and conduct much of a nation's innovative activity. Yet these smaller companies struggle in global markets on a far from level playing field, where large companies have distinct advantages.

To redress this imbalance the Commission saw it as a priority to improve the trading capability of the Small and Medium–sized Enterprises (SMEs), and perceived digital platforms as the modern means to this end. They considered that the best operational model for a vibrant Web2.0-based Internet services industry would be by analogy to well-performing biological ecosystems.

Open Source Software is adopted in the DBE[1]/OPAALS[2] projects as the best support for sustainability of such complex electronic webs, since it minimises interoperability problems, enables code access for cheaper in-house modification or development of systems, and reduces both capital and operating expenditure.

## 1. Origins and Development

### 1.1    Introduction

The best of biological communities have a population 'tail' comprising a large number of species, each present in small population numbers, many having niched out into specialised and demanding habitats, to which they have become peculiarly adapted. These exist alongside a small number of species present in very large population numbers; these are particularly favoured by the broadly prevalent environmental conditions of the time.

If changed conditions should wipe out the current dominants, then it is the tail that provides the best chance of saving the ecosystem from tipping into decline, since it harbours the genetic variance to respond to the new conditions[3]. Less healthy scenarios are where a dominant effectively produces a *monoculture*. If such a species is then taken out of the eco-equation, and there are no tail species to provide resilience to changed conditions, the ecosystem may fail.

---

[1] The **D**igital **B**usiness **E**cosystems Project: see the project book at www.digital-ecosystems.org/book. EU Sixth Framework Programme (FP6) - € 11 million. The project had 21 partners, including the London School of Economics and Political Science, Imperial College London, Trinity College Dublin (Ireland), Istituto Superiore Universitario di Formazione Interdisciplinare (Italy), IBM (Belgium), Intel (Ireland), Sun Microsystems (Spain), et al.

[2] The OPAALS project: **O**pen **P**hilosophies for **A**ssociative **A**utopoietic digita**L** ecosystem**S** at www.opaals.org. EU FP6 - € 9 million. Partners include the University of Cambridge, London School of Economics and Political Science, Food and Agriculture Organisation (UN), Tampere University of Technology (Finland), Universität Kassel (Germany), Indian Institute of Technology Kanpur (India), Ministry of Science and Technology (Brazil), National University of Ireland, National University of Rwanda, et al.

For industrial communities, if one accepts the analogy to biological ecosystems, the presence of super-dominants[4] can be regarded as a configuration that impacts all other 'business species'. Digital Business Ecosystems (DBE) and its sister project OPAALS are major initiatives which, together with their satellite work groups, focus on the health of the European SMEs – the 'business tail' – in the new knowledge-based economy that we are experiencing growing up around us.

A key observation is that when living in a marketplace alongside dominant firms, self-organisation by other firms is more difficult to achieve. The current Internet use models fail to support local autonomy for the firm [Dini 2008], and the so-called sharing/cooperation zones on the internet are homogenised in the direction of the interests of the dominant participants. Thus, conformity rather than diversity results[5].

This effectively centralised model further promotes the possibility of *single points of failure*[6], and more importantly *single points of control* from centralised servers. So an ecosystem-oriented approach necessitates a fully distributed peer-to-peer[7] networking model [Briscoe 2007b], which would avoid *single points of control* and so be immune to *single points of failure*, as shown in Figure 1.

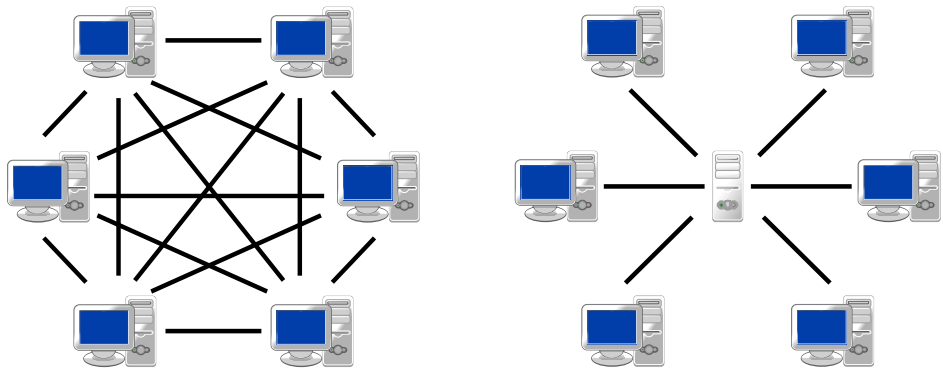

*Figure 1: Networking models: LEFT- A peer-to-peer system of nodes without central infrastructure. RIGHT - Centralised server-based service model.*

---

[3] Supported by biological theories of *founder* and *relic* populations, which may respond quickly to selection pressure.

[4] Large firms with greater than 70% of market share have been described as super-dominants in the European Court competition law cases.

[5] The 2009 US Hospitality eBusiness Strategies Report makes these statements in regard to online booking intermediaries, www.4hoteliers.com/4hots_fshw.php?mwi=3653:

     Total costs to hotels of as much as 30-40% from the booked hotel revenue [is taken].

     The Online Travel Agencies' merchant model has done long term damage to the hotel's brand and price integrity.

This latter statement refers to the homogenisation of room units with elimination of unique selling points such as location and special amenities which hoteliers and their clients value, together with the tendency to advertise special discount rates as standard rates.

[6] A Single Point of Failure is a part of a system which, if it fails, will stop the entire system from working. Such failures are undesirable in any system whose goal is high availability, be it a network, software application or other industrial system. [Dooley 2002]

[7] A peer-to-peer network, commonly abbreviated to P2P, is any distributed network architecture composed of participants that make a portion of their resources (such as processing power, disk storage or network bandwidth) directly available to other network participants, without the need for central coordination instances (such as servers or stable hosts). Peers are both suppliers and consumers of resources, in contrast to the traditional client-server model where only servers supply, and clients consume. [Schollmeier 2001]

## 1.2 Christening the DBE

The term Digital Business Ecosystems was constructed by adding 'digital' [Nachira 2002] in front of 'business ecosystem' [Moore 1996], to refer to a socio-economic development catalysed by Information and Communications Technologies (ICTs). Francesco Nachira emphasised the co-evolution of business ecosystems with their digital representations; hence '*Digital Business Ecosystems'* [Nachira 2007].

The term Digital Business Ecosystems had been used earlier than Nachira's casting of it, but with a focus exclusively on developing countries [Moore 2003]. The generalisation of the term under the EC model interpretation emphasised the *co-evolution* of the business ecosystem stratum and its complementary digital stratum.

Inspiration from ecosystems was sought because nature has been in the research business for 3.8 billion years and in that time has accumulated close to 30 million well-adjusted solutions to a plethora of design challenges that humankind struggles to address with mixed results [Benyus 2002]. *Bio-mimicry* is a discipline that seeks solutions by emulating nature's designs and processes, and there is considerable opportunity to learn elegant solutions for human-made problems [Benyus 2002]. Ecosystems are thought to be robust, scalable architectures that can automatically solve complex, dynamic problems, possessing several properties that may be useful in socio-economic technical systems, including self-organisation, self-management, sustainability and scalability [Levin 1998]. So, Digital Ecosystems can be defined as distributed adaptive open socio-technical systems, with properties of self-organisation, scalability and sustainability, inspired by natural ecosystems [Briscoe 2009b].

## 1.3 The Motivation and Incentive for Developing DBEs in the European Union

Both the Lisbon and Gothenburg strategies placed competitiveness firmly at the centre of political attention, and underlined the importance of creating a climate favourable to SMEs and the need to stimulate entrepreneurial initiative to achieve economic growth and sustainable development [Nachira 2007]. The Commission and the Council[8] also emphasised the requirement to use Open Source and Open Standards software in both business models and technical solutions. The whole DBE scheme as originally conceived was intimately linked to the aspirations of the European Research Area (ERA), and expected to feed results into the research programmes.

The DBE began life under the stimulus of the two Digital Divides[9] identified in Europe towards the end of the 1990s. First, it seemed clear that SMEs[10] in Europe were disadvantaged as compared to the larger firms; smaller companies were failing to make use of digital technologies at nearly all levels of functionality[11]. Second, a geographical schism seemed to have opened up between northern and southern Europe, with Nordic and Western European states broadly embracing new technologies, whilst southern and eastern Member States lagged behind. Urgency for action was precipitated by the increasing expansion of the Knowledge Economy.

---

[8] European Council held in Lisbon [23/24th March 2000]. The Heads of State asked the Council and Commission to draw up an action plan to meet the need (Brussels [13/03/2001], *– COM(2001)136 final.* The Commission's response was to launch two action plans: 'eEurope 2002', and 'eEurope 2005'. (See 28.5.2002 2002 COM(2002) 263, *eEurope 2005: an information Society for all: An Action Plan to be presented in view of the Sevilla European Council, 21/22 June 2002).* This is called the 'Lisbon Strategy'.

[9] See Eurostat's survey of 13 EU Member States plus Norway: gross sample was 100,000 enterprises; period [Nov 2000] to [June 2001], all sectors. (Conducted by Eurostat and the National Statistics Institutes. Sponsored by DG Enterprise.). The divide was most noticeable in e-business integration and associated skills.

[10] SME: small to medium-sized enterprise, that is 10 – 249 employees, designated the 'SME10s' in [Nachira 2002]. Francesco Nachira, (European Commission DG INFSO) headed the Digital Ecosystems Sector within the ICT (Information and Communication Technology) for Business Unit.

[11] In 2000, 16% of EU enterprises used the Internet for electronic delivery and 7% for e-payments; 18% of SME10s and 34% of large enterprises used e-ordering. Only 3% of EU enterprises had used Internet for e-commerce for more than 2 years. [20.02.2002] Eurostat, reported in Focus newsletter.

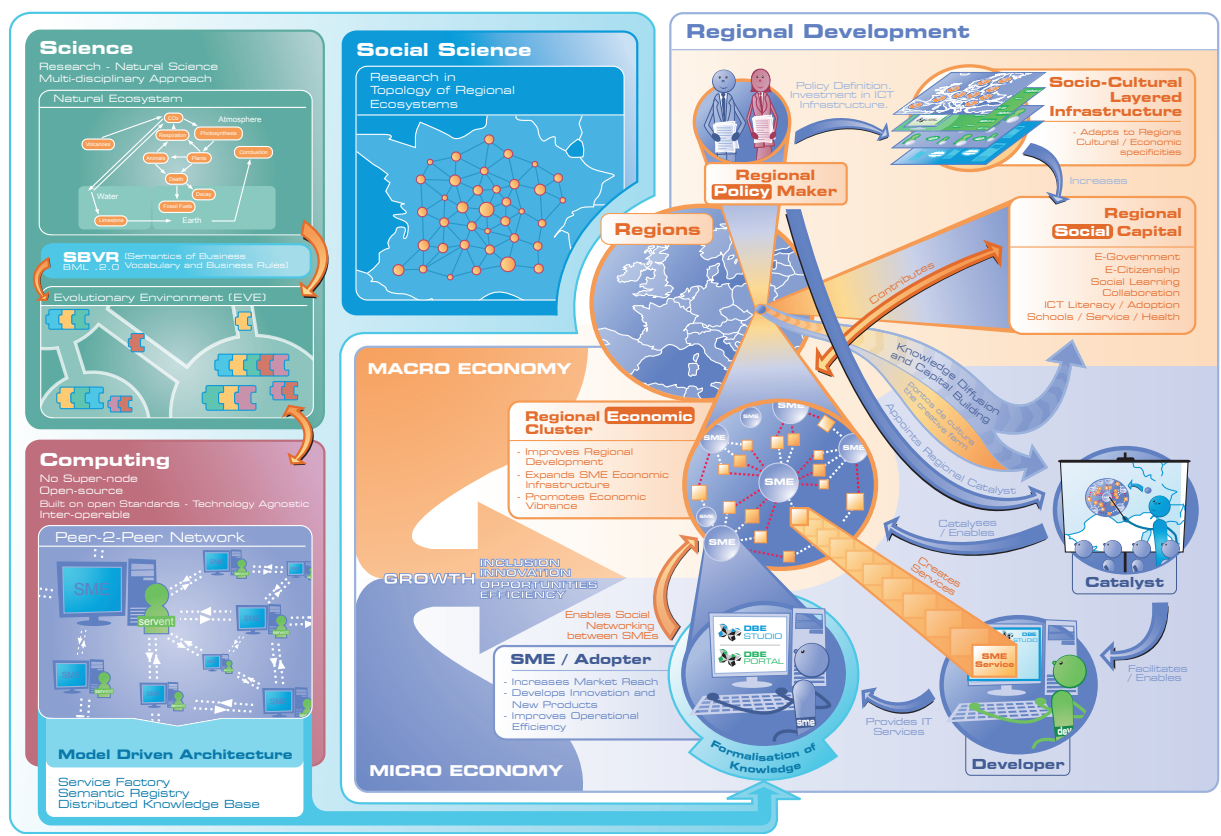

*Figure 2: An overview of the scope of the DBE project and its interdisciplinary nature.*

The European Union has strong overarching political and social agendas; it respects the survival of the small businesses as a social goal:

> *"The Community shall have as its task, by establishing a common market and economic and monetary union and by implementing the common policies or activities … to promote throughout the Community a harmonious and balanced development of economic activities, sustainable and non-inflationary growth respecting the environment."*[12]

Its key unit for funding is the *region*. Under the DBE initiative less fortunate regions would receive more aid[13], indicated by this statement promoting:

> *"a high degree of **convergence of economic employment**, a high degree of employment and social protection, the raising of the standard of living and quality of life, and **economic and social cohesion and solidarity among member states**"*[14]

The goal of the DBE project was to accelerate the progress of lagging regions and equalise their capacities to compete compared to leading regions.

The European Commission's answer to the imbalance in digital deployment was to construct an environment in which SMEs could thrive even against major players in a global market. They were to be strategically positioned on a network where synergies amongst them, notably in their supply chains

---

[12] Article 2 of the Treaty.

[13] Provided that this can be done 'without distorting competition'. See *eEurope 2002: An Information Society for all: An Action Plan prepared by the Council and the European Commission presented for the Feira European Council,* 19/20 June 2000.
Note that FP6 is now drawing to a close, and FP7 is in place to take over in the near future. See the Cordis website: http://cordis.europa.eu/fp7/home_en.html. Each of these framework programmes is a logical extension of the previous one.

[14] Article 2 of the Treaty.

and marketing, could be fostered. To succeed, such networks would need to be grounded in a platform having:

- Independence from a specific provider
- Open data formats and protocols, and code accessibility
- An infrastructure that satisfies a specific licensing model
- Usability and maintenance not depending on the goodwill of suppliers (notably if the supplier has a monopoly)
- Ability to replace services as more adequate ones appear (i.e. not be locked in).

Since the DBE concept aims to help local economic actors become active players in globalisation [Dini 2008], valorising[15] their local culture and vocations, enabling them to interact and create value networks at the global level, it has been embraced by the mayors and decision-makers of thousands of municipalities across Europe[16]. This approach, dubbed *glocalisation*, is being considered a successful strategy of globalisation that preserves regional growth and identity [Robertson 1994; Swyngedouw 1992; Khondker 2004]. It further serves to mitigate or exploit any possible tension between globalisation and localisation when adopting ICTs [Castells 2000].

## 1.4 An Overview of the Digital Ecosystem Research Cluster

**OPAALS**[17], a network of excellence (NoE) acts as the philosophical and theoretical grounding project of Digital Ecosystems research, tasked to establish and foster the idea of collaborative community software platforms for social and business purposes. The **DBE** project's role has been organising the strategic and tactical operations of mobilising software companies throughout the Union to launch a disruptive technology paradigm that could improve the SME value network across all sectors.

A group of satellite projects, such as **ONE**, **CONTRACT**, **Legal-IST** and **LEKTOR** variously investigated topics such as decentralised negotiation environments to satisfy the requirements of increasingly complex, global contractual agreements. This is important since current B2B electronic marketplaces and Internet trading platforms are centrally managed, not fully trusted and/or too expensive, thus not widely used by European SMEs [Telesca 2007]. Yet these SMEs must create tactical and strategic alliances in order to pursue business opportunities and growth. SMEs need to find trustworthy partners; but unfortunately access to reputation information is not readily available.

Legal-IST[18] explored Information Society Technologies (IST) - related legal support for SMEs. Its output is a road-map for an implementation based on the collected suggestions by European and national regulatory frameworks, with a target audience which includes the EC, policy-makers, consumers, trade and industrial organisations. LEKTOR[19], a Specific Support Action (SSA) project, spent its time raising awareness of potential legal obstacles to conducting e-Business.

Use of the new digital environments has the goal of reducing both transaction costs and time to market. Work done at one research institution[20] focused on developing methods to dynamically establish and manage contracts at runtime, and to apply formal verification techniques to collections of contracts in a Digital Ecosystem environment. The work also targeted applying monitoring techniques to contract implementation to help increase confidence levels in the business infrastructures.

---

[15] to valorize: to give or ascribe value or validity to an enterprise.

[16] See The Glocal Forum: A foundation in the field of city-to-city cooperation, which emphasize the central role of cities in international relations through their *Glocalization* vision. The Glocalization Manifesto:
www.glocalforum.org/mediagallery/mediaDownload.php?mm=/warehouse/documents/the_glocalization_manifesto.pdf

[17] The OPAALS project has 20 partners as far apart Brazil, India, the UK and Eire, and mainland Europe. Part-financed for four years by the European Union's 6th Framework Programme, OPAALS launched on 1st June 2006.

[18] www.ve-forum.org/apps/pub.asp?Q=1276&T=Clusters%20and%20Projects

[19] www.ubique.org/lektor

[20] www.ist-contract.org: a collaboration between one major industrial partner, three Universities, a Research Institute and three SME associates.

Moving closer to the regions and potential users, the **SEAMLESS** project, worked with the construction and textile sectors in an environment development called SEEM (Single European Electronic Market). The plan is to set up e-Registries across the member states with assistance of Chambers of Commerce, RDAs and entrepreneurial agencies. The aspiration is to create a marketplace without cultural or technological constraints[21].

Similar goals have inspired others. The **E-NVISION** project[22], which ended at the close of 2008 and focused on SMEs in the construction sector developing applications that would incorporate legal, economic, social and cultural services, facilitating their participation in future European e-business. **SATINE** had similar goals to E-NVISION targeting the tourism sector. It succeeded in enriching currently used Web Services Registries such as UDDI and ebXML with mechanisms to store and access Web Service semantics enabling easier discovery and automated composition of complex web services. The P2P network ethos of these projects permits facilitated discovery performing semantic routing of queries[23].

Understanding the vital importance of tool and die making to industrial supply networks **TOOL-EAST**[24] surveyed the sector in Eastern Europe and found it to be composed predominantly of SMEs without either financial or human resources to implement complex Enterprise Resource Planning (ERP) applications from powerful software suppliers. Further, the functionalities of these standardised applications fail when it comes to the specific requirements of the sector. The aim was to build a cost-efficient ERP application for tool and die making, freely adaptable under Open Source licensing. This allows co-ordination of intra-enterprise ordering, work planning and aims at support for e-collaboration in terms of process and data standards. Since demands of business software in other sectors with specific SME structures are highly comparable results from this work ought to be transferable.

The **VISP** project[25] also addressed the needs of collaborative enterprise amongst a cluster of SMEs via internet service provision (ISP). The work included a specification of ISP services by combining building blocks from an available list of a few hundred, each of which can be parameterised opening the possibility of providing tailoring and fine control, so distinguishing the product from that of the incumbent operators. Under the VISP scheme technical workflows will act on network components through abstract object representation based on existing standards such as MIB and CIM. Global data will be stored in LDAP following the Directory Enabled Networks (DEN) principles to be accessed by the cluster's partners.

A key strategy throughout these projects is to deploy *existing* valuable components via their integration at a higher level of abstraction.

The integration of all this work, and a study on rendering it accessible to policy-makers was undertaken by the **Peardrop** project[26] run by the Association Regionale Europeenne sur La Societe de L'information. The goal is to stimulate adoption of the DE way of doing e-Business. Here in the DE research cluster are tools, which are custom-made to support SMEs having marginal resources.

**EFFORT**[27], which closed at close of 2008, concentrated on the behaviour, governance, sustainability and constituency drivers of SMEs operating across regions and borders, examining legal frameworks under which they live. They focused on the probable needs of *virtual cluster* configurations. **EPRI-START** aimed at recruitment of suitable participants from the new member states of the Union to populate a *Qualified Partner Pool* for the IST programme.

--------------------

[21] The results of this project include six eRegistries established in Poland, Slovenia, Spain, Slovakia and Romania.

[22] www.e-nvision.org: a consortium of RTD organisations, a University, ICT consultants, Sees and heterogeneous clusters of organisations from 5 European countries.

[23] Work conducted at The Middle East Technical University Software Research and Development Center.: www.srdc.metu.edu.tr/webpage/projects/satine

[24] www.tooleast.org

[25] www.visp-project.org

[26] www.peardrop.eu

[27] www.effortproject.eu

Finally REgions for Digital Ecosystems Network (REDEN), is supported by projects such as the Digital Ecosystems Network of regions for (4) DissEmination and Knowledge Deployment (DEN4DEK). This thematic network aims to share experiences and disseminate all the necessary knowledge that will allow regions to plan an effective deployment of Digital Ecosystems at all levels (economic, social, technical and political) to produce real impacts in the economic activities of European regions through the improvement of SME business environments.

The Future Internet Enterprise Systems (FInES) cluster[28] will in future inherit/absorb the knowledge developed from the DBE and OPAALS initiatives and the satellite projects, and construct its vision of the future of the Internet from this and many external inputs.

## 1.5    The Anatomy of DBEs

We normally regard the DBE model to comprise two layers: the business stratum is a network – can be a cluster- of SMEs, the digital stratum reflects the relationships between the SMEs and other organisations and supplies technical support to facilitate their transactions.

### 1.5.1      The Business Stratum

The *business ecosystem* is an economic community supported by a foundation of interacting organisations and individuals; i.e. the *organisms* of the business world. This economic community produces goods and services of value to customers, who are themselves members of the ecosystem [Moore 1996]. A *healthy ecosystem* reflects a balance between co-operation and competition in a dynamic free market. Two main interpretations of its structure have been discussed in the literature.

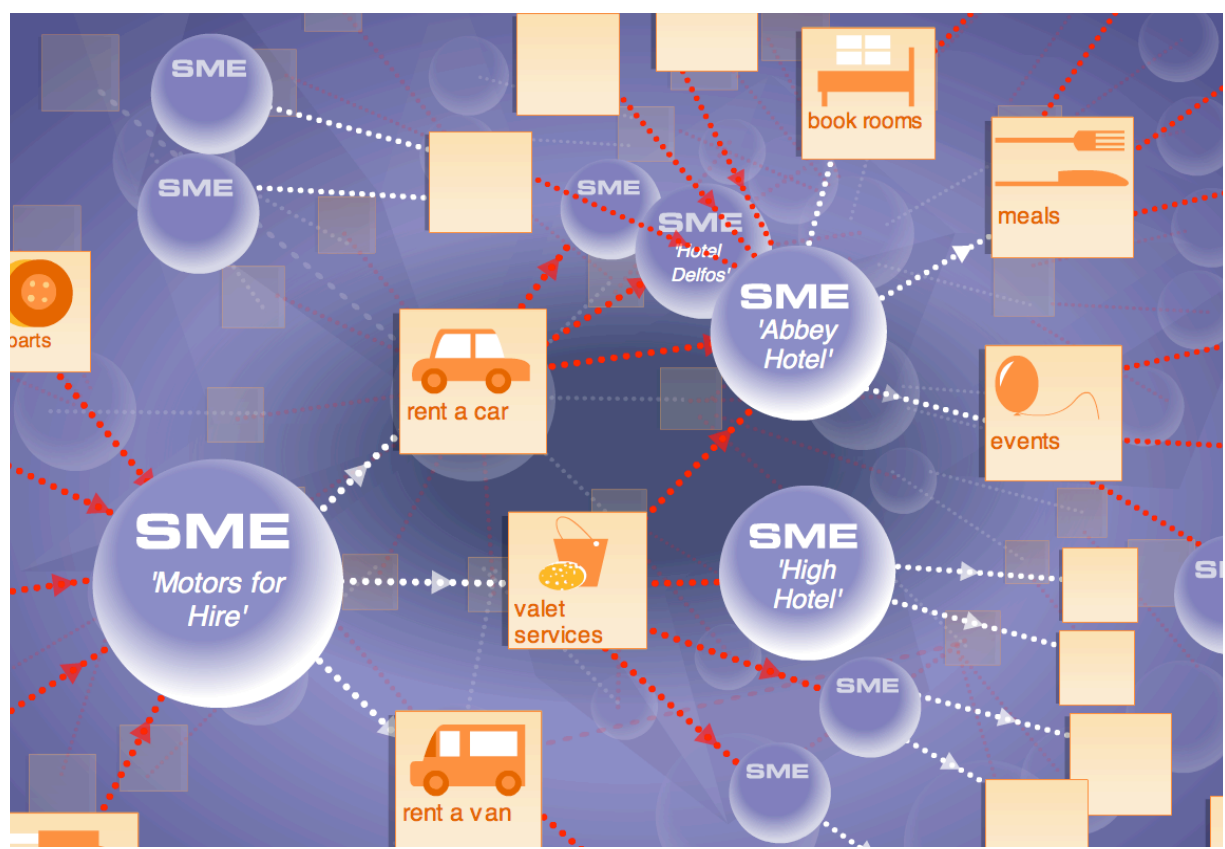

*Figure 3: The Business Ecosystem: [Nachira 2007]: Conceptual visualisation [English 2007] showing the interaction of Small and Medium sized Enterprise users, linked via the services they provide and consume. So is created a network of business ecosystems distributed over different geographical regions, business domains, and industry sectors.*

---

First, *the keystone model* has a structure in which a business ecosystem is dominated by a large firm surrounded by many small suppliers [Iansiti 2004]. This model works well when the central firm is healthy, but represents a significant weakness for the economy of the region when the dominant economic actor experiences difficulties [Moore 1996]. This model fits the economic structure of the USA, where a pattern of large enterprises positioned centrally to far-reaching value networks of suppliers is the norm [Iansiti 2004].

Second, the model for a business ecosystem developed in Europe shows properties of less *structure* and more *dynamism*; being composed mainly of SMEs, but able to accommodate large firms as well [Schmiemann 2006], as shown in Figure 3. All actors complement one another, leading to a more dynamic division of labour, organised along one-dimensional value chains and two-dimensional value networks [Corallo 2007]. This model is particularly well-adapted for the service and knowledge industries, where it is easier for small firms to reinvent themselves than, for instance, in the automotive industry which is dominated by large enterprises [Nachira 2007].

### 1.5.2 The Digital Stratum

A Digital Ecosystem is essentially the distribution of desired server functionality amongst a population of nodes[29]. Thus the resources provided by the nodes can be shaped into a *virtual data centre*[30] providing a platform for *Software-as-a-Service*[31] (SaaS). So *logically* we apprehend a single functionality, but *physically* that functionality is distributed across many machines.

While straightforward in principle, this strategy poses challenges on many different levels. An architecture for a Digital Ecosystem can be divided into layers, dealing with these challenges iteratively, since each level manages distinct concerns [Briscoe 2009]. The most fundamental layer, the Coordination Layer, deals with distributing coordination, which is taken for granted in homogeneous data centres where good connectivity, constant presence and centralised infrastructure can be assumed. One layer above, the Resource Layer, in which resource provision and consumption are arranged on top of the Coordination Layer. Whereas this is easily achieved in the homogeneous grid of a data centre, where all nodes have the same interests, it is more challenging in a distributed, heterogeneous environment. This middle layer would offer the usage experience of resource provisioning on the Platform-as-a-Service (PaaS)[32] layer and above. Including Utility Computing[33] scenarios, such access to raw storage and computation. Finally, the Service Layer is where resources are combined into end-user-accessible services, which may then themselves be further composed into higher-level services. So, access to services would be akin to a SaaS layer [Marinos 2009]. The layers relationships to one another are mapped in Figure 3. The Open System Interconnection (OSI) reference model is provided here as a contextual frame. The Digital Ecosystem is housed within the *application* layer. The diagram also shows the necessary technical infrastructure, legal framework and political support to provide required for the development and deployment of Digital Business Ecosystems.

---

[29] In a *node and edge* diagram, **nodes** represent digital systems or even single workstations, which in turn represent using institutions or single users. **Edges** represent communication links between nodes, which can be wired or wireless.

[30] A *data centre* is a facility, with the necessary security devices and environmental systems (e.g. air conditioning and fire suppression), for housing a *server farm*, a collection of computer servers that can accomplish server needs far beyond the capability of one machine [Arregoces 2003].

[31] Software-as-a-Service is a model of software deployment whereby a provider licenses an application to customers for use as a service on demand. SaaS software vendors may host the application on their own web servers or upload the application to the consumer device, disabling it after use or after the on-demand contract expires.

[32] Platform as a Service (PaaS) deliver a computing platform as a service. It facilitates deployment of applications without the cost and complexity of buying and managing the underlying hardware and software layers.

[33] Utility Computing is the packaging of computing resources, such as computation and storage, as a metered service similar to a traditional public utility (such as electricity, water, or the telephone network) [Rappa 2004].

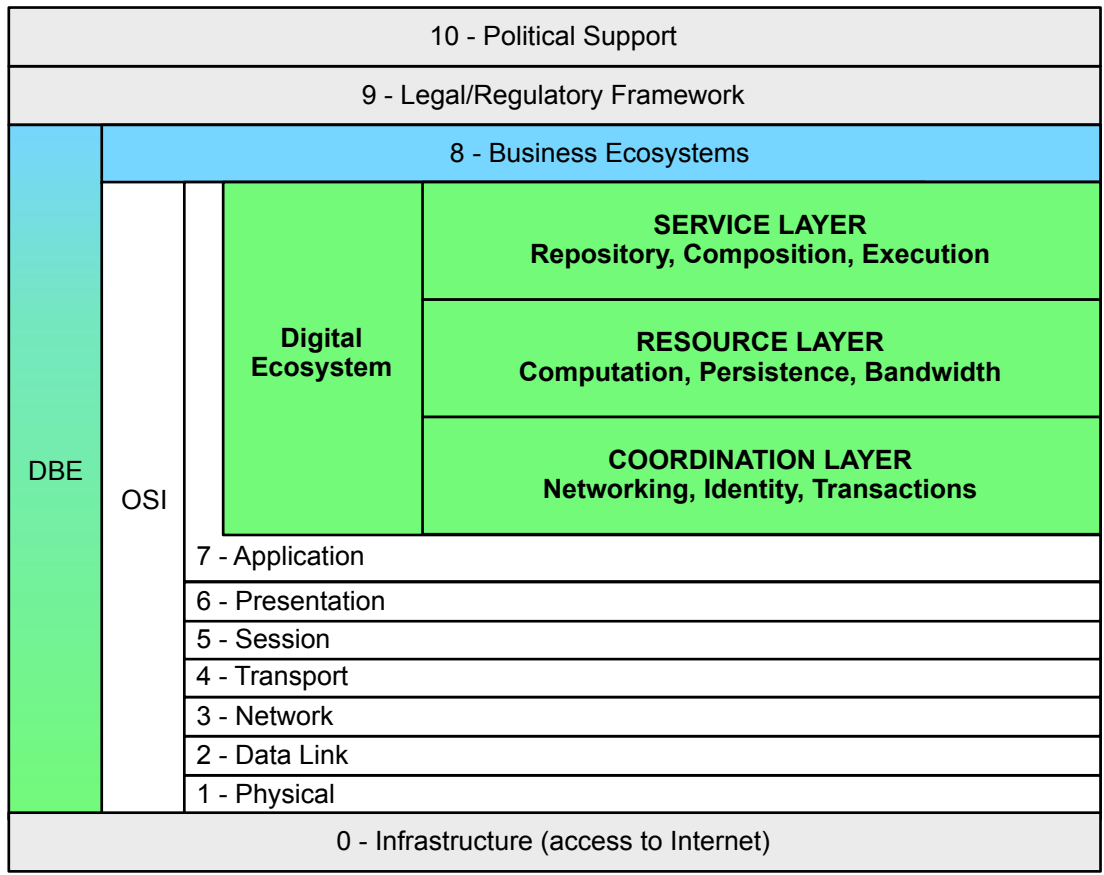

*Figure 4: The Digital Ecosystem: Its layers are shown in the wider context of its interactions with Business Ecosystems, including the Open System Interconnection (OSI) reference model as a reference point.*

We can see that the DBE represents business-to-business (B2B) interactivity supported by a software platform, which should have the desirable properties of a *natural ecosystem* [Briscoe 2007, Dini 2007], and follow a *Service-Oriented Architecture* (SOA) approach that adheres to the original SOA principles[34] [Razavi 2007; Nachira 2007**]**. Armed with this infrastructure the ecosystem can achieve evolution, self-organisation, and a self-optimising environment built upon an underlying *SOA* [Nachira 2007].

B2B interactions may be strongly influenced and motivated by social phenomena such as *small-world networks*[35], family ties and geographical proximity. In some cases, the need to rely on centralised third-party platforms can limit the formation, range and complexity of new value chains, business collaborations and business transactions [Dini 2008].

For B2B transactions across European countries (cross-border), it is also clear that reliable *trust systems* and an adequate legal and regulatory framework are needed [Dini 2008]. The characteristics of B2B transactions suggest the need for a distributed, open infrastructure that is interoperable and allows enterprises to move freely in the market, thereby avoiding lock-in from *principal agent problems*[36], which may arise from market failures.

---

[34] The following principles should guide the development, maintenance, and usage of SOAs: reuse, granularity, modularity, composability, componentisation, interoperability, and standards-compliance.

[35] They have many strongly connected clusters, called sub-networks, with a few connections between these clusters. Networks with this topology have a very high clustering coefficient and small characteristic path lengths. [Watts 1998]

The development and deployment of Digital Ecosystem-compliant architectures requires considerable support, beyond the vital technical innovations. These include the necessary infrastructural, legal and political support, which will be discussed in the following section.

### 1.5.2.1 Coordination Layer

A preferred means of coordination in the digital stratum would be to deploy a ***fully*** distributed system, because it would prevent vendor lock-in and control, issues of ever-increasing concern, through *distributed transactions* (i.e. without a third party observing)*,* trust-based identity (i.e. without depending on a third party) and maintain the *information privacy* of storage. In particularly sensitive cases of SMEs and start-ups, the provider-consumer relationship fostered between the owners of resources and their users could potentially be detrimental, as there is a conflict of interest for the providers. They profit by providing resources to up and coming players, but also wish to maintain dominant positions in their consumer-facing industries [Briscoe 2009]. For example, Google Apps for e-mail typically provides higher uptime than in-house e-mail servers [Montgomery 2008], but its failure [Perez 2007] highlighted the issue of lock-in that comes from depending on vendors. The even greater concern is the loss of information privacy, with vendors sometimes having full access to the resources they provide.

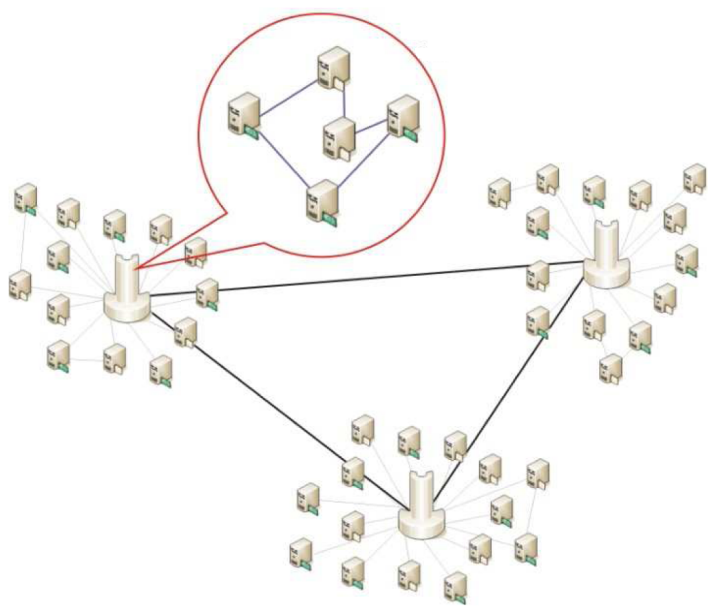

*Figure 5: Dynamic Virtual Super-peer [Razavi 2009]: An aggregation of machines with different qualities (e.g. in different time zones), which together create a stable virtual super-peer (e.g. available 24 hours a day). The aggregation of the machines is dynamic, changing when necessary, to maintain the stability of the virtual super-peer.*

The Coordination Layer could be achieved by a network of Dynamic Virtual Super-Peers (DVSPs) [Razavi 2008; Razavi 2009], as shown in the figure below. In order to understand the dynamic virtual super-peer[37] concept we need to appreciate the idea of a *virtual server* formed by *a collection of peers* (nodes on the network which represent institutions such as Universities, Research Centres and companies), and which *cooperate* to provide the functionality of *a single server* dynamically (i.e. when appointed to the role as a result of a majority vote given by all peers [Dini 2008]. Servers in this DVSP network would be *elected* to become members of the virtual super-peer cluster under criteria related to (1) their *availability* to share resources, and (2) their recent history of *reliability*. By the same token, any peer can be downgraded to a *regular* peer if its reliability or availability declines [Dini 2008].

---

[36] The principal agent problem in economics arises under conditions of incomplete and asymmetric information when a 'principal' hires an 'agent'. For example, the implementation of legislation (such as laws and executive directives) is open to bureaucratic interpretation, which creates opportunities and incentives for the bureaucrat-as-agent to deviate from the intentions or preferences of the legislators. Variance in the intensity of legislative oversight also serves to increase principal–agent problems in implementing legislative preferences. Various mechanisms may be used to try to align the interests of the agent with those of the principal, causing the agent to become 'locked in' to the principal. [Grossman 1983].

DVSPs would create a virtual network overlay on top of the existing physical network infrastructure. This would allow a peer to exchange messages with any other peers independent of network location (firewalls, NATs[38] or non-IP networks). It would also allow SMEs to communicate without needing to understand or manage complex and changing physical network topologies, and in addition allow mobile peers to move transparently from one location to another.

In distributed systems with variable node reliability, historical context is logically required to have certainty of node interactions. Fundamental to this context is the ability to identify nodes and therefore reference previous interactions. However, current identification schemes have *identity providers* [39]controlling provision.

The DBE will avoid *authority-centric* centralised identity provisioning (CIP) [Miyata 2006], in favour of *user-centric* [Maler 2008] distributed identity provisioning (DIP) [McLaughlin 2009], in which identity provision will be based on estimations of trustworthiness derived through social networks [McLaughlin 20010]. The Web 2.0 phenomenon [O'Reilly 2007] has shown the potential and possibility for identity provisioning through trust in social networks [Andersen 2007], i.e. *trust networks*. So, the DBE will involve *extending Web 2.0* social networking to distributed trust-based [Jøsang 2006] self-provisioning of identity through social networks.

DIP will also be a fundamental first step in creating dynamic virtual organisations (VOs) [Desanctis 1999] of SMEs aiming to compete with established large keystone firms [Iansiti 2004]. Such VOs can makes use of DIP, in which all the members share identity provisioning tasks and activities, but may equally choose to adopt internal or external (non-competing) CIP (benevolent dictatorship approach [Johnson 2004]). However, *proportional representation* schemes [Johnson 2004] would be possible, in which some SMEs of a VO retain full control over their identity provisioning activities. Therefore, fundamentally, going beyond just DIP, affording users choice over the social and power structures of their identity provisioning, rather than having CIP thrust upon them.

A key element of distributed coordination is the ability of nodes to jointly participate in transactions that influence their individual state [Marinos 2009]. Proprietary (centralised) solutions for transaction management often lack interoperability, i.e. if a SME uses one solution provider they cannot interact (transact) with an SME who happens to use another solution provider [Dini 2008]. DBEs will provide support for B2B transactions between SMEs[40] in a fully distributed way (no central point of control for transaction operations), needing to be highly resistant to fragmentation[41] as this may inhibit collaborative business interactions.

The approach of DBEs will rely on *local coordinators*, one for each service, that act independently of one another to guarantee (local) consistency and recovery support across all the transactions in a complex workflow representing a business process. Thus DBEs will be able to ensure local autonomy and loose coupling because they rely on distributed rather than on centralised transaction management. Ideally, appropriately annotated business processes could be executed over a distributed network with a transactional model maintaining the ACID[42] properties usually considered fundamental on behalf of the initiator; other directions of research include relaxing these properties to maximise concurrency [Vogel 2008].

---

[37] Many P2P systems use stronger peers (super-peers, super-nodes) as servers and client-peers are connected in a star-like fashion to a single super-peer.

[38] In computer networking, network address translation (NAT) is the process of modifying network address information in datagram (IP) packet headers while in transit across a traffic routing device for the purpose of re-mapping a given address space into another. Most often today, NAT is used in conjunction with network masquerading (or IP masquerading) which is a technique that hides an entire address space, usually consisting of private network addresses, behind a single IP address in another, often public address space.

[39] Users, typically, have one identity per service, and the *service provider* (e.g. Google, Yahoo!) provides the identity, via an internal identity provider (IdP). This IdP is responsible for retaining information pertaining to this identity, providing authentication and authorisation, and, *generally speaking,* presiding over the entire life-cycle of the identity. [McLaughlin 2009, McLaughlin 2009b]

[40] A B2B transaction between SMEs may involve simple usage of a web service or composition of several services from various service providers. A business transaction may be finished over a period of minutes, hours, or even days – thus the term, from a computational perspective, of long-lived or long-running transaction.

[41] A situation where the network gets divided into smaller isolated networks

### 1.5.2.2 Resource Layer

As the networking infrastructure is now in place from the previous layer, we can discuss the first consumer-facing uses of the *virtual data centre*. First, distributed computation, which has a long history of successful *centrally controlled* incarnations (e.g. SETI@home[43]). However, the DBE will need to provide distributed coordination of the computational capabilities that nodes offer through the Coordination Layer [Briscoe 2009].

Next, distributed persistence (storage) will naturally be required on participating nodes. However, the method of information storage in the DBE is an issue with multiple aspects. First, information can be file-based or structured. Second, while constant and instant availability can be crucial, there are scenarios in which recall times can be more relaxed. Such varying requirements call for a combination of approaches, including distributed storage [Yianilos 2001], distributed databases [Garciamolina 2008] and key-value stores[44] [Bain 2008]. Information privacy would be provided by the encryption of user information when on remote nodes, only being unencrypted when cached on the user's node, allowing for the secure and distributed storage of information.

Finally, bandwidth management via peer-to-peer protocols such as BitTorrent[45], which have made the distribution of information over networks much less bandwidth-intensive for providers, accomplished by using the downloading peers as repeaters of the information they receive. The DBE would have to make use of such approaches to ensure the efficient use of available network bandwidth, to avoid fluctuations or sudden rises in demand burdening parts of the network.

### 1.5.2.3 Service Layer

DBEs represent a new era for Service-Oriented Computing[46] (SOC), making services explicitly dependent on other resource providers instead of building on self-sufficient resource locations. DBEs make this property more explicit, breaking down the *stand-alone service paradigm* by having any service, by default, composed of resources contributed by multiple participants.

DBEs aim to support network-based economies reliant on next-generation ICT, extending the (SOA) concept with the automatic combining of available and applicable services in a scalable architecture, to meet business user requests for applications that facilitate business processes [Briscoe 2006b].

The vision of SOAs is the assembling of application components from a loosely coupled network of services, that can create dynamic business processes and agile applications that span organisations and computing platforms. The great promise of SOA is that the *marginal cost* of creating the n-th application is virtually zero, as all the software required *already exists* and can satisfy the requirements of other applications. The interaction of services (all of which are hosted by unassociated peers) is *specified by users* in an ad-hoc way, with the intent driven by newly emergent business requirements [Leymann 2002]. Unfortunately, the current generation of SOA implementations for addressing B2B requirements tend to underestimate the negative impact of the new unique proprietary functional models. There are complex mechanisms that motivate ICT and business communities to adopt some standards that may become *de facto* standards. *Good* standards do not always emerge from competition, and current implementations of SOA are unsatisfactory because they can be shown to violate important principles.

---

[42] In computer science, ACID (atomicity, consistency, isolation, durability) is a set of properties that guarantee that database transactions are processed reliably.

[43] SETI@home is a distributed computing (grid computing) project using Internet-connected computers, hosted by the Space Sciences Laboratory, at the University of California, Berkeley, to analyse radio signals, searching for signs of extra terrestrial intelligence.

[44] A distributed storage system for structured data that focuses on scalability, at the expense of the other benefits of relational databases [Bain 2008], e.g. Google's BigTable and Amazon's SimpleDB.

[45] BitTorrent is a peer-to-peer file sharing protocol used for distributing large amounts of data. BitTorrent is one of the most common protocols for transferring large files.

[46] Service-Oriented Computing promotes assembling application components into a loosely coupled network of services, to create flexible, dynamic business processes and agile applications that span organisations and computing platforms [Papazoglou 2003]. This is best achieved through a Service-Oriented Architecture (SOA), an architectural style that guides all aspects of creating and using business processes throughout their life-cycle, packaged as services.

We can now see that the service layers of DBEs must be compliant with the principles of SOC. It must also include a *Distributed Service Repository (DSR)* that provides persistence (storage), as with traditional service repositories, for the pointers to services and their semantic descriptions. To support the absence of service-producing nodes during service execution, i.e. provider SMEs being offline, there must also be persistence of the executable code for services. Naturally, the implementation of a distributed service repository is made easier by the availability of the distributed storage service infrastructure.

When a service is required, but is not currently instantiated on a suitable node, *a copy* should be retrieved from the DSR and instantiated as necessary, allowing for flexible responsiveness and resilience to unpredictable traffic spikes. Remote service execution would need to be secured against potentially compromised nodes, through trust networks and perhaps through encrypted processing schemes [Gentry 2009]. Since delivering a service over large distances in the network comes at a potentially high cost, the lack of a central 'well-connected server' calls for a fundamental paradigm shift, from *pull*-oriented approaches to hybrid *push/pull*-oriented approaches [Briscoe 2009b, Briscoe 2006].

While some arguments would seem to advocate an extension of the *distributed* principle to the rest of the OSI application layer and beyond, what we are actually advocating is *choice*. For example, at this stage it is not considered necessary to require a fully distributed solution (alternative) to the Domain Name System (DNS)[47], which while nominally distributed, remains under centralised control both technologically and organisationally, permitting numerous distortions in the network. These include *domain squatting*[48], abuses by domain registrars [Lyon 2009], subjection to political control [Mueller 2002], and risks of the infrastructure being compromised [Goodin 2008]. Also, in terms of the required infrastructure, should reliance on ISPs become to problematic, alternative forms of *mesh networking*[49] could be considered and advocated.

## 2. Dominants in a Marketplace

### 2.1 Introduction

In the introduction to section 1 of this paper we described the types of natural ecosystem and discussed their relative robustness. We subsequently demonstrated that centralised systems had drawbacks that distributed systems could mitigate by careful design of infra-structures on principles such as those used for the DBE. We noted that the presence of a dominant in a given habitat set limits on the prospects of smaller species.

---

[47] The Domain Name System (DNS) is a hierarchical naming-space for computers, services, and other resources participating in the Internet. It translates domain names meaningful to humans into their counterpart numerical identifiers associated with networking equipment to locate and address these devices world-wide [Mockapetris 1988]. So, translating human-friendly computer host-names into Internet Protocol (IP) addresses, e.g. www.example.com translates to 208.77.188.166.

[48] Domain squatting (also known as cybersquatting) is registering, trafficking in, or using a domain name in bad faith, with the intent to profit from the goodwill of a trademark belonging to someone else. The cybersquatter then offers to sell the domain to the person or company who owns a trademark contained within the name at an inflated price [Maury 2002].

[49] Mesh networking is a type of networking wherein each node in the network may act as an independent router, regardless of whether it is connected to another network or not. It allows for continuous connections and reconfiguration around broken or blocked paths by "hopping" from node to node until the destination is reached. [Akyildiz 2005].

In this section we describe economic patterns of dominance in software markets, firstly characterising proprietary software platforms, and secondly, exploring products from what is loosely called 'the gift economy' as sustainable alternatives.

We also describe the peculiar market phenomenon – network effects - that occurs when the good in question is a shared resource such as a file format or software platform. We ask the question: could a software platform qualify as an *essential facility*? If so, specific legal constraints can be applied to ensure global access to it.

Dominant software publishers operate in markets characterised as simultaneously servicing and being part of the knowledge economy. Their products embody intellectual property that typically represents large upfront, sunk costs for R&D. Marginal costs thereafter, although not zero, are relatively low, owing to the means of replication and distribution of software[50]. This early-day investment typically leads to a felt need to protect, via Intellectual Property Rights (IPRs), expensively achieved R&D results. Defended IPRs, however, may set up substantial barriers to entry of other market participants.

The IP issue is further complicated by the asymmetry of objectives amongst the types of IP protection available. Patent supposedly represents a clear-blue-water *difference* (usually an improvement) in the *idea* for a software product. However, software developments tend to be incremental rather than step advances, so grants of patent (theoretically) might be expected to be few and far between. In addition, the reliability of searches of prior art (notably in the US PTO[51]) does not lend confidence that either the *novelty* or *non-obviousness* criteria for patentability can be readily satisfied. If they are not satisfied, then the patent is vulnerable to validity challenges.

Copyright weakly protects the idea for a functionality inherent in computer code via its *expression*. Under copyright law only the *expression* of an idea, and not the idea itself, is protected. The would-be copyist of an innovation embedded in low level computer code is frustrated by having to reverse analyse (engineer) the code to access *its idea*. However, once attained, the idea can be cast into a different expression (differently coded) to appear legitimately as a separate product. In any event, *copyright stands for no test of true technological advance.* It springs into being immediately the software is *fixed* (ie recorded in some medium); its threshold of originality amounts to mere *origination* by the author who claims it, rather than representing a striking innovation.

Trade secret, in the context of computer programs, denies release of the program's source code, distributing only the executable level.

## 2.2 Dominance and Monopoly

'If a monopoly firm's profits are to persist in the long run, effective entry barriers must prevent the entry of new firms into the industry'[52].

---

[50] First comer advantages are an important factor since either this, or hard work can lead to market dominance.

[51] Patent and Trademark Office.

[52] Lipsey and Chrystal Princs of Economics Ed 9, at 165.

In a monopoly[53], one entity (or a cartel[54] entity in lockstep) produces and sells a product with no close substitutes[55]. Economically the monopolist's demand curve slopes down, implying that the price must drop to attract more customers[56].

The definition of a monopoly is widely accepted to stand or fall by the definition of '*the relevant market*'[57]:

> the *outer bounds* of a product market are determined by the reasonable interchangeability of use or cross-elasticity of demand between the product itself and substitutes for it.

'Monopoly power' can be demonstrated circumstantially by a conjunction of two elements: a dominant market share in a <u>well-defined</u> relevant market[58] AND the existence of *significant barriers of entry* to that market, which we examine more closely in the next section.

---

[53] Areeda *et al* define monopoly as (1) the 'ability to price above the competitive level and (2) to persist in doing so for a significant period without erosion by new entry or expansion' [Areeda 1995].

[54] Small number of colluding producers/sellers (an oligopoly) that together monopolise the market.

[55] Judge Penfield Jackson in *US v Microsoft Corporation,* (US District Court for the District of Columbia. Civil actions no 98-1232. [Nov 1999]) concluded that there were no close substitutes for the Windows operating system. See the cross-examination of John Romano and Frank Santos (Hewlett Packard), Bart Brown (Gateway), Stephen Decker (Compaq) to select the testimony of just a few.

[56] A classical view of the implications of a monopoly is this: monopolies reduce consumer welfare in that (1) income redistribution from consumer to monopolist is pure profit. (In social welfare terms these redistributed amounts offset each other to produce neutrality) (2) deadweight loss (the cost to society of a market that does not work optimally) is transferred <u>away</u> from consumers <u>yet not captured</u> by producers. The net result is that fewer than optimal units are produced, and fewer resources are allocated to production of the good than perfect competition would require.

[57] In US case law, that definition stems from *US v du Pont de Nemours* (the 'Cellophane case'), 351 US 377 [1956]; and *Brown Shoe v US,* 370 US 294 [1962].

[58] The US Department of Justice *Merger Guidelines* make this widely-used definition:

> A Market is defined as product or group of products and the geographical area in which it is produced or sold, such that an hypothetical profit-maximising firm, not subject to price regulation, that was the only present and future producer or seller of those products in that area likely would impose at least 'a small but significant and non-transitory increase in price', (SSNIP) above the competitive level, assuming the terms of sale of all other products held constant. (See DOJ Guidelines at P27 s1.32 '*Firms That Participate*').

> A Relevant Market - See also P18/19 Guidelines - for the test of smallest group of products where the monopolist will <u>profitably</u> impose at least 'a small but significant and non-transitory' increase in price. This is got by iteratively applying the test to the expanded product group until substitutes run out.

In Microsoft the *Findings of Fact* (*US v Microsoft Corporation*, US District Court for the District of Columbia. Civil actions no 98-1232. [Nov 1999]). identified these relevant markets:

(1) The licensing of all Intel-compatible PC operating systems world-wide.
(2) Web browsing functionality.

## 2.3    Network Effects

Software platform markets display marked *network effects*. Not only the direct customer, or 'joiner' effect, but also indirect effects via the 'shadow' networks of distributors, application writers, trainers and so forth that are closely, vertically, related to the primary network.

In such markets an *installed base*[59] belonging to a dominant first-comer supplier in the market has *value* to consumers over and above the intrinsic value of the supplied item[60]. This is the case in, for example the operating system market, where clusters of software applications have grown up around the prevalent market dominant operating system. The gravity pull of a dominant platform is so intense that independent software vendors (ISVs) have little choice but to write to APIs of the platform on which nearly all software will run[61]. This makes it difficult for other operating system producers (open source or proprietary) to enter the market and compete. That is, <u>*applications barriers to entry*</u> against new competitors are erected, and consumers become locked in to the product of the vendor they began with. Costs of 'consumer switch' [to an entrant's product] may be high. If the number of applications written for the entrant's platform is sparse (almost certain to be the case in the early reaches of the entrant's project) the consumer is severely punished for the switch.

A monopolist supplying into the market will likely handle the market by:

> ➢ Price discrimination to harvest the early adopters prepared to offer the high initial price
> ➢ Location of new classes or sources of customers
> ➢ Renting not selling the product
> ➢ Upgrading the product, even planning in obsolescence.

The pattern can become one of upgrading on a very frequent basis, as is the case with much office worker software. This strategy can abolish backward compatibility between versions. Gejlsbjerg and Stanley [Gejlsbjerg 2005] reported this 'migration funnel' effect in the case of Microsoft's network server operating systems:

> When a [Windows] Domain having a legacy system is first installed with Windows 2000 it is in *mixed mode*. This can be changed to *native mode but the change is not reversible*. In native mode, Windows NT 4.0 Domain Controllers cannot participate in the Domain[62] and *therefore a server that is only interoperable with Windows NT 4.0 OS, can no longer be used*[63]. The full advantage of any upgrade to Windows 2000 is therefore only felt when workgroup servers in the Domain (other than the Domain Controllers) are Windows 2000-compatible; for example they support Microsoft *Kerberos* [the Windows authentication system for users][64]. Microsoft therefore recommends:

> > To receive *maximum benefit* from Windows 2000 technologies and fully realize your migration-related goals, *it is recommended that you switch your Windows 2000 Domains to native mode as soon as possible*[65].

---

[59] Installed base: a substantial number of customers are set up with a good or service (a computer operating system, for example) that makes it easy for them to communicate with others having the same facility. Examples are: exchanging files that have the same format, and swapping applications that run on the same platform.

[60] *Network effects*, or *network externalities*, means this: a good's value to the buyer of an extra unit is higher when more units are sold, everything else being equal [Economides 2001].

[61] Since it may not be cost effective to write to more than one platform.

[62] See Step-by-Step Guide to Managing Active Directory (http://technet.microsoft.com/en-us/library/bb742437.aspx).

[63] See also para 170 of the EC Microsoft Decision.

[64] Saved in :\Windows\System32\Kerberos.DLL

[65] Microsoft, Deployment Planning Guide, at Chapter 10: Determining Domain Migration Strategy, Migration Goals. Available from: www.microsoft.com/windows/techinfo/reskit/dpg/default.asp

In two modern court decisions against the Corporation (US[66], EC[67]), the judgements were based on Microsoft's strategic market behaviour, components of which are implemented by technological means.

The Commission found that Microsoft then occupied a dominant position in the PC operating system market[68] under Article 82 of the Treaty[69], with the added ingredient that Microsoft controlled the *de facto* standard of that market, and had done so for some time[70], leaving room for only fringe competition. It is dominant in *the server operating system market*, which was found to be a global market by the US courts and assessed by market share[71]. Microsoft may be said to be super-dominant in the PC operating system market worldwide (over 90% market share[72]).

The market for PC operating systems was of interest in the US decision[73]. Both markets were addressed in the Commission Decision. Also objected to in the US case was the tying of a web-browser to Windows[74]. In the European decision Microsoft was held to have bundled a streaming media player[75] with Windows[76]).

---

[66] We focus on Judge Penfield Jackson's 'Findings of Fact'. Lopatka - Lopatka and Page: Economic Authority and the Limits of Expertise in AT Cases, 90 Cornell LR [2005] 619 - points to Judge Jackson's effort in Microsoft to set forth a distinct separate set of Findings of Fact (US v Microsoft Corporation, US District Court for the District of Columbia. Civil actions no 98-1232. [Nov 1999]). Other commentators intimate that Judge Jackson made strenuous efforts to publish the Findings before the impending Presidential Election, when a reversal of Democrat fortunes might impact activity in the FTC and DOJ. The Findings were later complemented by Judge Jackson's 'Conclusions of Law' (03 04 [2000] US District Court District Columbia, *US v Microsoft,* Civil Action No 98-1232 1232 (TPJ)). The factual findings have the reputation of never been substantively challenged.

[67] European Commission Decision C(2004)900 final, [24.03.2004], re Case COMP/C-3/37.792 Microsoft given by Mario Monti, and the ensuing judgement of the Court of First Instance, Vesterdorf's Order: Case compete T-201/04R.

[68] Commission Decision at (429).

[69] As expressed in *United Brands v Commission* [1978] (ECR 207, at para 65: 'a position of economic strength enjoyed by an undertaking (firm) which enables it to prevent effective competition being maintained on the relevant market by affording it the power to behave to an appreciable extent independently of its competitors, its customers and ultimately of the consumers'.

[70] Its market share was quoted in The Decision as 92.1% in terms of unit shipments; 92.8% in terms of revenue, rising to 93.8%, and 96.1% respectively in [2002]. Coming from 76.4% in [1997].

[71] The Commission explained the methodologies used to arrive at Microsoft's position in the server market in s5.2.2.1, beginning at (para 473-488). In [2002] of all servers shipped in the popular under-25,000-US$ range, Microsoft Windows' share was 64.9% of units shipped. Measured by revenues it was 61.0%. Fringe competitors were Netware: 9.4%, Linux: 13.4%, and UNIX: 11.1%. The latter figures per units shipped. (at para 491). (Source: IDC Worldwide Quarterly Server Tracker). Other figures from the Commission's own inquiries come in even higher (at para 495). The conclusion was that Microsoft holds a share of at least 50%, and for most measures 60-75%.

[72] See EC Decision at 1062. Commentators on the hi-tech market reflect that market share may simply mirror first comer advantage or it may signal the hard-fought winning of the industry standard.

[73] This account is based strongly on the *Findings of Fact*. (Judgement delivered by Judge Penfield Jackson in the District Court District Columbia: US v MS. Civil action no 98-1232 (TPJ), and State NY v Microsoft Civil action no 98-1233 (TPJ), US District Court for the District of Columbia. 84 Suppl 2nd, 9 [1999]. Decided: [Nov 5th 1999]).

## 2.4    The Position of Minor Species in Software Markets

A recent survey of Very Small Businesses (VSBs) in the East of England [Stanley 2008] concluded that lack of interoperability amongst software versions and components generated the most difficulties for this class of business.

The findings were that digital systems adopted by a firm often decay at an alarming rate rather than stabilising in use. Replacement of components can cause interoperability failures. Combating this system erosion can drain time and other resource from the firm, which a VSB cannot easily absorb. These requirements for a better service offering were distilled in the report:

➢    The development of systems relatively immune to version incompatibilities
➢    Version changes at a lower frequency
➢    Stability across time for *instructions* to perform simple computing tasks, or at least some degree of instruction language genericity and continuity within product.
➢    Secure file-sharing facilities and a range of available collaborative digital platforms.
➢    Clarity in licensing provision
➢    Clear specifications from software publishers of *all system components required* for the level of tasking the company will need when it initially acquires the software, without a confusion of options leading to poor investment choices and subsequent additional expense.

Work done by a small London-based company in specifying and installing Open Source software has identified the third (charitable) sector as another class of organisation that finds that proprietary licensed software is heavily discounted (for example Microsoft products can be supplied at one tenth the market price. This is an attractive proposition, but less so when it is appreciated that should modifications be needed, the code is closed, and the licensee is beholden to the supplier for all changes since it lacks ownership of code[77]. Since the third sector is increasingly important as economy support, for example as member of a PFI, and members are frequently nationally based, they therefore need substantial ICT provision, and any adverse conditions for them in terms of costs or efficiency need to be addressed.

## 2.5    The Copyright Basis for Software Licenses

### The Perils of the 'EULA': End User License Agreement

Licensors may withhold the EULA until after purchase of the software or hide the EULA. On challenge in one US lawsuit licensors and retailers agreed to show the EULA before purchase. Encryption of software can make it impossible for the user to install their new software without accepting the EULA

---

[74] The earliest popular Web browser Navigator, a product of Netscape Communications Corp was first shipped in December [1994]. Microsoft introduced its Internet Explorer browser in July [1995].

[75] The Commission decided that since 'the market provides media players separately' (p 804), and since 'some operating system users *will not need or want* a streaming media player at all' (p 807) there was a separate market for browsers. The Commission admitted, however, that 'the direct consumer demand test' the Rolls Royce test of what constitutes 'a market' … *focuses on historic consumer behaviour, likely before software integration'* (p 808). The objectionable behaviour was to render it difficult to impossible for consumers to buy the OS without the player (at 826-834). The Commission concluded that tying has identifiable effects on the competition in the market for media players (at 835-954).

[76] Microsoft maintained that: 'multimedia playback '*functionality'* has been '*integrated'* in Windows since 1992', opposing the EC suggestion that a tying event instigated the tying in [1999], Microsoft's Statement of Objection. (Submission 17th Oct 2003 at 135 –138, cited Decision (para 1056)).
The Commission found Microsoft dominant in the OS market (at 799) and that Streaming Media players and OSs are two separate products (at 800-825).

[77] *Bind Technology (www.bindtechnology.com)* led by Jerome Josephra. (pers comm).

*or* violating the Digital Millennium Copyright Act (DMCA) or its foreign counterparts. (See *Blizzard v BnetD[78]*).

There can be errors with price lists that confuse the licensee. Scott Braden, writing in December 2005[79] tells how he uncovered a plethora of mistakes in the resale market for one major software publisher's products.

He counted 43,000 unique items in one major proprietor's list, with confusion over the UIDs amongst these items. This would necessitate clients performing their own calculations to remain compliant with their license. For example, on an instalment plan, suppose a 3-year contract costs $1,066. Divide the price by 3 and expect to pay $355.33 for the first instalment. You will need to be quite sure of the years remaining on your license since this may be unclear in the documentation. In certain cases the license duration – irrespective of whether it be for one, two, or three years – has the same identifier. Finally, you must remember the due date in case your reseller forgets the license renewal date.

The EULA began life as a liability disclaimer and contract between the user and the publisher of the software. However, terms of a EULA are all too frequently not thoroughly read and understood the user. This next discussion reviews some of the risks the user public runs by casually signing a EULA.

1. Agreements '*not to criticise the product publicly*' effectively prevents watchdogs from making independent quality surveys. Thus information about products disseminated to consumers is less complete than it should be.

2. Similarly if the customer agrees not to disclose the results of any benchmark test on the software to any third party, this inhibits unfettered, informed choice and hence fair competition.

3. If a EULA contains a term such as '*using this product means you will be monitored*', automatic license updates can be coerced by having the networked device contact some third party without the knowledge of the consumer. This jeopardises security and threatens privacy. Some EULAs provide for the licensor to charge the user's credit card an 'automatic' fee when her subscription runs out.

4. Certain Digital Rights Management (DRM) clauses enable downloads onto the user's computer. One example reserves the licensor's right to download 'security-related updates' which 'may impair the software', without liability.

5. It is perfectly legal in US law to reverse engineer products provided that this is a prescribed 'fair use'[80] and provided the result is a non-infringing product[81]. Yet the clause '*do not reverse engineer this product*' appears in some EULAs. Annalee Newitz suggests that this stifles innovation and reinforces lock-in[82]. She maintains that such a clause also inhibits customisation of the software. Napster users must agree, amongst other things not to 'reverse engineer or *to emulate the functionality [of the software]*'. Such clauses are problematic. Is it reasonable to expect a user to know the exact bounds of *legal* reverse engineering? The EULA may use some form of words such as '*except and only to the extent that such activity is expressly permitted by applicable law*'. This opaque expression gives the user no guidance as to how a violation could occur, yet it obeys the strict legal requirement to give him notice.

6. Licensors can receive a fee to have advertising software bundled with licensed products. In signing the EULA the end user may also be promising not to remove this ride-along software, and not to use sniffer (detector) programs to access or interrupt traffic from this software on the wire. The data so collected has obvious marketing value.

7. There may be clauses to ensure that the user *has also* signed up to future modified versions of the EULA. This implies a one-sided modification of a contract without one party's explicit consent.

---

[78] United States Court of Appeals, (Eighth Circuit). Case no: 04-3654 [2005].

[79] *How a Mid-sized Company Saved over $870,000 on a $3M Microsoft EA* [specific kind of license agreement] *in Less than Three Weeks*. Braden claims to have corrected, as he puts it: 'mistakes that cost my client several hundred thousand dollars over the past few years' (www.microsoft-watch.com/content/operating_systems/how_to_negotiate_with_microsoft.html).

[80] 17 US Code, (the Copyright title).

[81] *Sega v Accolade*. Appeal in 9th Circuit, 977 F.2d 1510, [1992].

[82] www.eff.org/wp/dangerous-terms-users-guide-eulas

8. The EULA may contain disclaimers for destruction of the functionality of the user's existing software by the new licensed programs. Such disclaimers may clash with the consumer's legal rights.

## 2.6    Attractions of Adoption of DBEs

Open Source software for ICT provision is attractive for the following reasons:

There are no license costs and consequently no need to audit software licenses to determine whether they are within the period and other license conditions. The argument implies that manpower need not be deflected from core business activity to housekeeping. The OS[83] property of 'stability across time' mitigates the requirement to cope with compatibility issues or retraining.

Both these factors ensure that ICT provision is sustainable, and costs are more predictable. This ability to plan costs on a long-term basis is an asset to VSBs and third sector entities and would be of immense value to the public sector (public purse) in the economic downturn.

OS ICT systems could be set up as local agent-based for installation and maintenance. Many Universities across UK have been early users of Unix (first wave OSS), and early adopters of Linux operating systems (second wave). As a consequence graduates from such institutions can provide in-house support, and graduate employment opportunities could be created in the process. This is to be compared with expensive Help Desk support on which proprietary systems rely. In the OS world, effectively the community (which developed the software) becomes the Help Desk. The beauty of this last scenario is its vista of a pool of *community-based* and *community-controlled* know-how, harbouring transferable skills *local* to the region, yet supported by a large *global* know-how base.

The concept of 'the region' may, for economic purposes, be regarded as *a principality*. We know that in 2003 - 04 the 3.8 million small businesses in the UK spent close to £17 billion on information technology and telecommunications[84]. We further have the report from the same source that about half these businesses intended to increase their IT budgets in 2004. So for the East of England, for example, we might estimate how much money is currently leaving the region in the form of software licenses which, being renewable, are not providing optimal sustainability, in fact they are producing a drain of resources in sectors least able to manage the loss.

We are aware of the Gordon Brown 2004 *prudential borrowing scheme[85]*. The scheme allowed councils to borrow money (repayable out of future income or savings). So, for example, Newcastle could borrow £20 million to implement the plan laid out in an in-house bid [for a replacement software system]. Over an 11½-year period, £13 million was to be spent on 'technical refresh' (replacing all the council's servers and computers on a rolling cycle) and £7 million on new systems and implementation. Educational sector reports indicate costs of refresh undertakings[86].

Yet despite the advantages indicated for Open Source provision, as mandated in the DBE scheme, there remain barriers to more widespread adoption of Open Source, community-based systems, as we see in the next section.

### 2.6.1    Barriers to Adoption for the SMEs

The Commission identified a set of obstacles to full Internet use by the SMEs. Much the same reasoning applies to the adoption of a fully formed Open-source-based digital ecosystem, yet the list contains the very reasons why a DE would be desirable. These include:

---

[83] **O**pen **S**ource software.

[84] Source: Microsoft booklet, *Selling Microsoft Solutions*, [2004].

[85] Speech by Gordon Brown to the Local Government Association conference in Harrogate: http://www.hm-treasury.gov.uk/press_80_03.htm

[86] Becta report, published Jan 2009 - Microsoft Vista and Office 2007: Final report with recommendations on adoption, deployment and interoperability:
http://publications.becta.org.uk/download.cfm?resID=35275

➢ Lack of technical knowledge and expertise amongst SMEs[87]

➢ Lack of technological solutions and issues of interoperability[88]

➢ The costs of setting up the e-business architecture, deflecting capital from the firm's core business[89].

➢ Legal/regulatory mismatch when trading digitally across borders[90].

## 2.7    A Balance Between Protecting Truly Innovative Software and Sustainable Platforms

Bruce Perens[91]  [Perens 2005] separates the roles of Open Source and closed code. All firms need enabling software platforms: operating systems, email, office applications, and other services to run their core businesses. This he terms *non-differentiating software*. This is the infrastructure of the firm and its surrounding ecosystem. Such software does not define the core business nor <u>does it</u> offer *specific support* to that core, that is the role of *differentiating software* that defines the firm and its Unique Selling Point (USP) and supports *only* the firm's unique core business.

Whereas the former ought to be as network-compatible and open standard as possible to enable the flow of information into, through and out of the firm, the latter must be allowed to be closed for commercial security reasons. Both strategies bespeak their own rewards.

Work towards the DE project includes a study of the high technology cluster in Cambridge UK. It is no coincidence that this University town, renowned for its history of innovation in hard science also boasts an *investor cluster* committed to foster and support the University spin-out and independent start-up companies within the range of the '*Cambridge Technopole*'. It is a *sine qua non* that investors in cutting edge technologies will need IPRs from the firm in order to attract their money. The platforms that support the UK economy, however, need to shed closed code and closed standards in order to allow the recovering UK knowledge economy to flourish in the second decade of the millennium.

## Conclusions

In an old market-based economy, made up of sellers and buyers, the parties exchange property. In a new network-based economy, made up of servers and clients in a *business ecosystem*, the parties share access to services and experiences. Digital Business Ecosystems are a platform for a network-based economy of *business ecosystems*, providing the necessary technical infrastructure and legal mechanisms for the creation of networked economies.

The DBE project began development of a *reference implementation*[92]  for the Digital Business Ecosystems platform, which once completed will be part of the regional deployment of Digital Ecosystems in DEN4DEK and OPAALS. So, the next major step in our research will be to collect real world data, confirming that Digital Ecosystems operate effectively with business ecosystems in creating Digital Business Ecosystems.

---

[87] The Commission pointed to (1) a serious shortage of skilled ICT professionals (estimated in 2002 to be 1.9M, projected to rise to 3.8M in 2003), (2) the importing of skills from other countries, and (3) the bringing into the firm of costly contract experts, all of which practices it finds unsatisfactory. (Nachira at 5).

[88] The Commission report suggested that the SMEs, above all other firms, have an interest in 'fully compatible, <u>open</u>, interoperable ICT solutions that stay relatively stable across time', and hence an avoidance of the lock-in problem.

[89]  The EC report amplified: there is, for example, the worry of making an *<u>initial expensive mistake</u>* and *<u>the mounting costs of maintenance</u>*. Obstacle removal suggested by the Commission would be the use of open source software, together with open and distributed common infrastructure. (Report at 19).

[90] Which factor the DBE legal projects sought to address.

[91] Perens is a co-founder of open source, specifically of Linux Standard Base. Perens wrote the Debian Social Contract, a portion of which later became *The Open Source Definition*.

[92] In the software development process, a reference implementation is the standard from which all other implementations, with their respective customisations, are measured, and to which all improvements are added.

**Acknowledgements**

This work was supported by the European Commission through the OPAALS Network of Excellence. We would also like to thank Jerome Josephra of Bind Technology (www.bindtechnology.com) for useful suggestions.

---